# Common Voting Rules as Maximum Likelihood Estimators[*]

**Vincent Conitzer**
Computer Science Department
Carnegie Mellon University
5000 Forbes Avenue
Pittsburgh, PA 15213
conitzer@cs.cmu.edu

**Tuomas Sandholm**
Computer Science Department
Carnegie Mellon University
5000 Forbes Avenue
Pittsburgh, PA 15213
sandholm@cs.cmu.edu

## Abstract

Voting is a very general method of preference aggregation. A voting rule takes as input every voter's vote (typically, a ranking of the alternatives), and produces as output either just the winning alternative or a ranking of the alternatives. One potential view of voting is the following. There exists a "correct" outcome (winner/ranking), and each voter's vote corresponds to a noisy perception of this correct outcome. If we are given the noise model, then for any vector of votes, we can compute the maximum likelihood estimate of the correct outcome. This maximum likelihood estimate constitutes a voting rule. In this paper, we ask the following question: *For which common voting rules does there exist a noise model such that the rule is the maximum likelihood estimate for that noise model?* We require that the votes are drawn independently given the correct outcome (we show that without this restriction, all voting rules have the property). We study the question both for the case where outcomes are winners and for the case where outcomes are rankings. In either case, only some of the common voting rules have the property. Moreover, the sets of rules that satisfy the property are incomparable between the two cases (satisfying the property in the one case does not imply satisfying it in the other case).

## 1 Introduction

Voting is a very general method for aggregating multiple agents' preferences over a set of alternatives, such as potential presidents, joint plans, allocations of goods or resources, *etc.*. As such, it is a topic of significant and growing interest in the AI community with applications in collaborative filtering [13], planning among automated agents [10, 11], determining the importance of web pages [1], to name a few. Recent AI research has studied the complexity of executing voting rules [7], the complexity of manipulating elections [3, 2, 5], as well as efficient elicitation of the voters' preferences [4, 6]. In this paper, we study how voting can be interpreted as a maximum likelihood estimation problem. We believe this will shed new light on, and enhance the applicability of, both voting and maximum likelihood techniques.

To see how voting may be interpreted as a maximum likelihood problem, let us first contrast the following two views of voting:

1. The voters' preferences over candidates are idiosyncratic, and there is no sense in trying to model where they came from. The purpose of voting is solely to find a compromise candidate that maximizes the combined welfare of the agents, either explicitly or implictly trading off agents' utilities against each other.

2. There is some absolute sense in which some candidates are better than others, which is prior to and not dependent on the agents' preferences. Rather, the agents' preferences are merely their noisy estimates of this absolute quality. The purpose of voting is to infer the candidates' absolute goodness based on the agents' noisy signals, *i.e.*, their votes.

For the purposes of this paper, we will be concerned with the second interpretation (which is especially sensible in contexts such as document selection). It is instructive to visualize this interpretation as the simple Bayesian network given below. In this network, the observed variable is the agents' votes, and we seek to estimate the correct outcome. The most natural way of doing so is to take the maximum likelihood estimate of the correct outcome.[1] Which outcome is chosen as the maximum likelihood estimate (MLE) depends on the conditional probability table for the "agents' votes" node—that is, it depends on the noise model. Each noise model generates a maximum likelihood estimator function from agents' votes to outcomes, and this function constitutes a voting rule.

[*]This material is based upon work supported by the National Science Foundation under ITR grants IIS-0121678 and IIS-0427858, and a Sloan Fellowship.

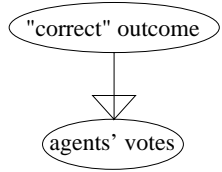

The basic idea of using an MLE approach to voting was introduced as early as the 18th century, by Condorcet [8]. Condorcet studied a particular noise model in which a voter ranked two candidates correctly with some given probability $p > 1/2$. Condorcet solved the cases of 2 and 3 candidates.[2] The solution for arbitrary numbers of candidates was given two centuries later by Young [15]: he showed that it coincided with a voting rule proposed by Kemeny [12]. A slightly extended model where $p$ is allowed to increase with the distance between two alternatives in the correct ranking of alternatives, and the rules that this produces, has also been studied [9]. However, none of these rules correspond to any of the commonly used voting rules. Should the fact that we do not know how to interpret the common voting rules as maximum likelihood estimators be viewed as a criticism of the common voting rules? This is perhaps somewhat unfair, because we may simply not yet have studied the right noise models to generate these rules. Nevertheless, there is no guarantee that such noise models exist. Moreover, it is useful to actually *know* a noise model for a rule: it increases our understanding of the rule, it allows us to question the assumptions in the noise model (and therefore in the rule), and, where we disagree with these assumptions, we can modify the noise model to obtain a rule that is more appropriate for our needs.

In this paper, we address these issues by turning the traditional approach on its head, by asking: *For which common voting rules does there exist a noise model such that the rule is the maximum likelihood rule for that distribution?* Perhaps surprisingly, if we assume that the voters' votes are independent given the correct outcome, it turns out that only some rules have this property.

Answering our question has at least the following purposes:

• Rules that have the property are in a sense more natural than ones that do not, especially in settings where information aggregation is the main purpose of voting. Hence, answering the question will provide some guidance in the problem of choosing a rule.

• Showing that a rule does have the property requires us to construct a noise model, for which the rule is the maximum likelihood estimator. Subsequently, we can assess whether the noise model is reasonable, or needs to be modified to become reasonable. In the latter case, the modified noise model may lead to a novel and useful voting rule.

The rest of this paper is organized as follows. In Section 2, we define the common voting rules and define the types of noise model that we consider. In Section 3, we present our positive results (noise models that have common voting rules as their maximum likelihood estimator). In Section 4, we present a technique for showing that *no* noise model has a given rule as its maximum likelihood estimator, and apply this technique to obtain our negative results.

## 2 Definitions

### 2.1 Voting rules

We have a set of *candidates* (aka. *alternatives*) $C$ over which the voters vote. A *vote* is defined as a strict ordering (ranking) of the candidates in $C$. A *(voting) rule* takes the (vector of) votes as input, and produces an *outcome*. This outcome can either be a single candidate (the winner), or a ranking of the candidates (where the top-ranked candidate is the winner). Most rules allow for the possibility of ties, and do not specify how ties should be broken. In this paper, for our positive results, we do not attempt to realize any particular tie-breaking rule; our negative results hold regardless of how ties are broken.

We next define the common voting rules that we study.

• *scoring rules*. Let $\vec{\alpha} = \langle \alpha_1, \ldots, \alpha_m \rangle$ be a vector of integers such that $\alpha_1 \geq \alpha_2 \ldots \geq \alpha_m$. For each voter, a candidate receives $\alpha_1$ points if it is ranked first by the voter, $\alpha_2$ if it is ranked second *etc*. The score $s_{\vec{\alpha}}$ of a candidate is the total number of points the candidate receives. The *Borda* rule is the scoring rule with $\vec{\alpha} = \langle m-1, m-2, \ldots, 1, 0 \rangle$. The *plurality* rule (aka. majority rule) is the scoring rule with $\vec{\alpha} =$ $\langle 1, 0, \ldots, 0, 0 \rangle$. The *veto* rule is the scoring rule with $\vec{\alpha} = \langle 1, 1, \ldots, 1, 0 \rangle$. Candidates are ranked by score.

---

[1] A Bayesian (maximum a posteriori) interpretation is, of course, also possible: Bayes' rule gives $P$(correct outcome|agents' votes) = $P$(correct outcome)$P$(agents' votes|correct outcome)/$P$(agents' votes). If the distribution over correct outcomes is uniform, this expression is maximized by the maximum likelihood estimate.

[2] This approach leads to inconsistent cyclical rankings (*e.g.* $a \succ b \succ c \succ a$) with nonzero probability, but this does not affect the maximum likelihood approach.

- *single transferable vote* (STV). The rule proceeds through a series of $m - 1$ rounds, each one eliminating one candidate. In each round, the candidate with the lowest plurality score (that is, the least number of voters ranking it first among the remaining candidates) is ranked at the bottom of the remaining candidates. Then, that candidate is eliminated from the votes (each of the votes for that candidate "transfer" to the next remaining candidate in the order given in that vote). The last remaining candidate is the winner.

- *Bucklin.* For any candidate $c$ and integer $l$, let $B(c, l)$ be the number of voters that rank candidate $c$ among the top $l$ candidates. Candidate $c$'s Bucklin score is $\min\{l : B(c, l) > n/2\}$, and candidates are ranked according to their scores (where lower scores are *better*). That is, if we say that a voter "approves" her top $l$ candidates, then we repeatedly increase $l$ by 1, and whenever a candidate becomes approved by more than half the voters ("passes the post"), that candidate is placed next in the ranking. When multiple candidates pass the post simultaneously, ties are broken by the number of votes by which the post is passed.

- *maximin* (aka. *Simpson*). Let $N(c_1, c_2)$ be the number of votes that rank candidate $c_1$ higher than candidate $c_2$. Candidate $c$'s maximin score is $\min_{c' \neq c} N(c, c')$ (the candidates worst score in a *pairwise election*). Candidates are ranked by score.

- *Copeland.* A candidate $c$ gains one Copeland point for every pairwise election it wins (one point for every $c'$ such that $N(c, c') > N(c', c)$), and loses one Copeland point for every pairwise election it loses (minus one point for every $c'$ such that $N(c, c') < N(c', c)$). Candidates are ranked by score.

- *ranked pairs.* Sort all ordered pairs of candidates $(a, b)$ by $N(a, b)$, the number of voters who prefer $a$ to $b$. Starting with the pair $(a, b)$ with the highest $N(a, b)$, we "lock in" the result of their pairwise election $(a \succ b)$. Then, we move to the next pair, and we lock the result of their pairwise election. We continue to lock every pairwise result that does not contradict the ordering $\succ$ established so far.

### 2.2 Noise models

In this paper, we will place the following restrictions on the noise model. First, we require that the noise is independent across votes. That is, votes are conditionally independent given the correct outcome, as illustrated by the Bayesian network below.

Second, we require that the conditional distribution given the correct outcome is the same for each vote

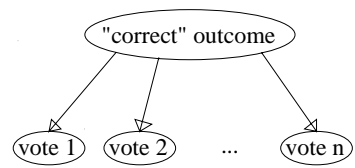

(for example, the first vote cannot be more likely to agree with the correct outcome than the second vote). Together, the two restrictions amount to the noise being i.i.d.

These restrictions strengthen our positive results that show that certain rules can be interpreted as maximum likelihood estimators. As for the negative results that certain rules cannot be so interpreted, it turns out that without any restrictions on the noise model, the question becomes trivial. Specifically, if we do not make the restriction that votes are drawn independently given the correct outcome, then *any* rule is an MLE, as the following trivial proposition shows.

**Proposition 1** *Any voting rule $\rho$ can be interpreted as a maximum likelihood estimator (if we do not require that votes are drawn independently given the correct outcome). Moreover, if the rule is anonymous (it treats all voters symmetrically), then the noise model can also be anonymous (*i.e. *the conditional distribution given the correct outcome is the same for each vote).*

**Proof**: The following noise model will suffice: given any correct outcome, let the probability of all vote vectors on which $\rho$ produces an outcome that is different from the correct outcome be 0, and let the probability on all other vote vectors be positive. We note that if the rule is anonymous, then this is an anonymous noise model. ∎

Hence, in the remainder of the paper, we make the above two restrictions. We say that a rule that can be viewed as a maximum likelihood estimator under these restrictions when the outcome is a winner is an *MLEWIV* (Maximum Likelihood Estimator for Winner under I.i.d. Votes) rule, and one that can be viewed as a maximum likelihood estimator under these restrictions when the outcome is a ranking is an *MLERIV* (Maximum Likelihood Estimator for Ranking under I.i.d. Votes) rule.

## 3 Voting rules that can be interpreted as MLEs

In this section, we lay out our positive results: we show which of the common voting rules can be interpreted as maximum likelihood estimators under the assumption of i.i.d. votes (both for the case where the outcome is the winner and the case where the outcome is a ranking).

It turns out that scoring rules satisfy both the MLEWIV and MLERIV properties.

**Theorem 1** *Any scoring rule is both an MLEWIV and an MLERIV rule.*

**Proof**: We first show that every scoring rule is an MLEWIV rule. If candidate $w$ is the winner in the correct outcome, then let the probability (given the correct outcome) of a vote $j$ that ranks candidate $w$ in position $r_j(w)$ be proportional to $2^{s(r_j(w))}$ (where $s(r)$ is the number of points a candidate derives from being ranked $r$th in a vote). Thus, the probability of votes 1 through $n$ given the correct outcome is proportional to $\prod_{j=1}^{n} 2^{s(r_j(w))} = 2^{\left(\sum_{j=1}^{n} s(r_j(w))\right)}$. Of course, $\sum_{j=1}^{n} s(r_j(w))$ is exactly candidate $w$'s score. Now, the problem of finding the maximum likelihood estimate of the correct outcome is the problem of choosing a candidate $w$ to maximize the former expression. This is done by choosing the candidate $c$ with the highest score $\sum_{j=1}^{n} s(r_j(c))$. Hence every scoring rule is an MLEWIV rule.

We next show that every scoring rule is also an MLERIV rule. If the candidates are ranked $c_1 \succ c_2 \succ \ldots \succ c_m$ in the correct outcome, then let the probability (given the correct outcome) of a vote $j$ that ranks candidate $c_i$ in position $r_j(c_i)$ be proportional to $\prod_{i=1}^{m} (m+1-i)^{s(r_j(c_i))}$. Thus, the probability of votes 1 through $n$ given the correct outcome is proportional to $\prod_{j=1}^{n} \prod_{i=1}^{m} (m+1-i)^{s(r_j(c_i))} = \prod_{i=1}^{m} (m+1-i)^{\left(\sum_{j=1}^{n} s(r_j(c_i))\right)}$.

Of course, $\sum_{j=1}^{n} s(r_j(c_i))$ is exactly candidate $c_i$'s score. Now, the problem of finding the maximum likelihood estimate of the correct outcome is the problem of labeling the candidates as $c_i$ so as to maximize the former expression. Because $m+1-i$ is positive and decreasing in $i$, this is done by labeling the candidate $c$ with the highest score $\sum_{j=1}^{n} s(r_j(c))$ as $c_1$ (thereby maximizing the number of factors $m+1-1 = m$ in the expression), the candidate with the second highest score as $c_2$, *etc*. Hence every scoring rule is an MLERIV rule. ∎

STV, on the other hand, satisfies only the MLERIV property (we will see later that it violates the MLEWIV property).

**Theorem 2** *The STV rule is an MLERIV rule.*

**Proof**: Let the candidates be ranked $c_1 \succ c_2 \succ \ldots \succ c_m$ in the correct outcome. Let $\delta_j(c_i) = 1$ if all the candidates that are ranked higher than $c_i$ in vote $j$ are contained in $\{c_{i+1}, c_{i+2}, \ldots, c_m\}$, and let $\delta_j(c_i) = 0$ otherwise. Then, let the probability of vote $j$ be proportional to $\prod_{i=1}^{m} k_i^{\delta_j(c_i)}$, where, for every $i$, $0 < k_i < 1$, and $k_{i+1}$ is much smaller than $k_i$. Thus, the probability of votes 1 through $n$ given the correct outcome is proportional to $\prod_{j=1}^{n} \prod_{i=1}^{m} k_i^{\delta_j(c_i)} = \prod_{i=1}^{m} k_i^{\left(\sum_{j=1}^{n} \delta_j(c_i)\right)}$.

Again, the problem of finding the maximum likelihood estimate of the correct outcome is the problem of labeling the candidates as $c_i$ so as to maximize this expression. To do so, we must first minimize the number of factors $k_m$, because these factors dominate. The number of such factors is $\sum_{j=1}^{n} \delta_j(c_m)$, which is the number of votes that rank the candidate that we label $c_m$ first. Hence, we must rank the candidate that is ranked first the fewest times last (as $c_m$). Next, we must minimize the number of factors $k_{m-1}$. The number of such factors is $\sum_{j=1}^{n} \delta_j(c_{m-1})$, which is the number of votes that rank the candidate that we label $c_{m-1}$ either first, or second after $c_m$. Hence, we must rank the candidate that is ranked first the fewest times *after $c_m$ is removed* second-to-last (as $c_{m-1}$)—*etc*. Thus the maximum likelihood estimate of the ranking is exactly the STV ranking. Hence, STV is an MLERIV rule. ∎

## 4 Voting rules that cannot be interpreted as MLEs

In this section, we will show that some rules cannot be interpreted as maximum likelihood estimators under the restriction that the votes are i.i.d. (given the correct outcome). To do so, we rely on the following lemma.

**Lemma 1** *For a given type of outcome (e.g. winner or ranking), if there exist vectors of votes $V_1$, $V_2$ such that rule $\rho$ produces the same outcome on $V_1$ and $V_2$, but a different outcome on $V_1 + V_2$ (the votes in $V_1$ and*

$V_2$ taken together), then $\rho$ is not a maximum likelihood estimator for that type of outcome under i.i.d. votes. Specifically, if there exist vectors of votes $V_1$, $V_2$ such that rule $\rho$ produces the same winner on $V_1$ and $V_2$, but a different winner on $V_1 + V_2$, then $\rho$ is not an MLEWIV rule; if there exist vectors of votes $V_1$, $V_2$ such that rule $\rho$ produces the same ranking on $V_1$ and $V_2$, but a different ranking on $V_1 + V_2$, then $\rho$ is not an MLERIV rule.[3]

**Proof**: Let $s$ be the outcome that $\rho$ produces on $V_1$ and on $V_2$. Given a distribution such that $s \in \arg\max_{s'} P(V_1|S = s')$ and $s \in \arg\max_{s'} P(V_2|S = s')$ (where $S$ is the correct outcome), we have $s \in \arg\max_{s'} P(V_1|S = s')P(V_2|S = s') = \arg\max_{s'} P(V_1 + V_2|S = s')$. But $s$ is not the outcome that $\rho$ produces on $V_1 + V_2$, so $\rho$ is not a maximum likelihood estimator for the distribution. ∎

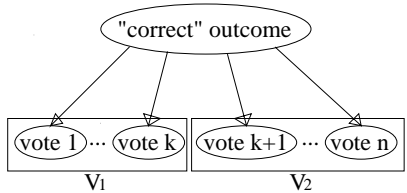

Figure 1: *Graphical illustration of Lemma 1. If the maximum likelihood estimate for the correct outcome given only the votes $V_1$ coincides with the the maximum likelihood estimate given only the votes $V_2$, then it must also coincide with the maximum likelihood estimate given all the votes $V_1 + V_2$.*

Below, when we use Lemma 1 to show that a rule is neither an MLEWIV nor an MLERIV rule, we will exhibit vectors of votes $V_1$ and $V_2$ such that the rule produces the same ranking on them (and hence also the same winner), but the rule produces a different winner on $V_1+V_2$ (and hence also a different ranking).[4]

**Theorem 3** *The Bucklin rule is neither an MLEWIV nor an MLERIV rule.*

**Proof**: We will apply Lemma 1 to both cases. Let $V_1$ contain two votes $a \succ b \succ c \succ d \succ e$, and one vote $b \succ a \succ c \succ d \succ e$. The following describes at which points the candidates pass the $n/2$ votes mark. $a$ is ranked the top candidate by two votes; $b$ is ranked among the top two candidates by all votes; $c$ is ranked among the top three candidates by all votes; and $d$ is ranked among the top four candidates by all votes. Hence the ranking produced by the Bucklin rule on $V_1$ is $a \succ b \succ c \succ d \succ e$.

---

[3]Such paradoxical outcomes are perhaps reminiscent of a known paradox in which statistical tests in two different subpopulations both suggest that a treatment is helpful, but when the data of the two tests are aggregated, this suggests that the treatment is actually harmful [14]!

[4]We note that a proof using Lemma 1 that a rule is not MLEWIV is not sufficient to show that the rule is not MLERIV, because in the proof, $V_1$ and $V_2$ may produce the same winner but different rankings.

Let $V_2$ contain two votes $b \succ d \succ a \succ c \succ e$, one vote $c \succ e \succ a \succ b \succ d$, and one vote $c \succ a \succ b \succ d \succ e$. The following describes at which points the candidates pass the $n/2$ votes mark. $a$ is ranked among the top three candidates by all votes; $b$ is ranked among the top three candidates by three votes; $c$ is ranked among the top four candidates by all votes; and $d$ is ranked among the top four candidates by three votes. Hence the ranking produced by the Bucklin rule on $V_2$ is $a \succ b \succ c \succ d \succ e$, the same as on $V_1$.

Now, consider the ranking that the Bucklin rule produces on $V_1 + V_2$. The following describes at which points the candidates pass the $n/2$ votes mark. $b$ is ranked among the top two candidates by five votes; $a$ is ranked among the top two candidates by four votes; $c$ is ranked among the top three candidates by five votes; and $d$ is ranked among the top four candidates by six votes. Hence the ranking produced by the Bucklin rule on $V_1 + V_2$ is $b \succ a \succ c \succ d \succ e$. ∎

We have already shown that STV is an MLERIV rule, which implies that the condition of Lemma 1 for an MLERIV rule does not hold. Still, it is interesting to see directly why this condition does not hold. Suppose that we have votes $V_1$ and votes $V_2$, on which STV produces the same ranking. The candidate $c_m$ ranked last by STV in both of $V_1$ and $V_2$ receives the lowest number of votes in both cases, and therefore must also receive the lowest number of votes in $V_1 + V_2$, and be ranked last in this case as well. Then, the candidate $c_{m-1}$ ranked second last in both of $V_1$ and $V_2$ receives the lowest number of votes in both cases after the removal of $c_m$, and therefore must also receive the lowest number of votes in $V_1+V_2$ after the removal of $c_m$, and be ranked second last in this case as well—*etc.* Hence the ranking produced by the STV rule on $V_1+V_2$ must agree with that produced on $V_1$ and $V_2$.

However, this does not yet imply that the Lemma fails on STV for the MLEWIV property, and in fact it does not:

**Theorem 4** *The STV rule is not an MLEWIV rule.*

**Proof**: We will apply Lemma 1. Let $V_1$ contain three votes $c \succ a \succ b$, four votes $a \succ b \succ c$, and six votes

$b \succ a \succ c$. Given these votes, $c$ drops out first; its three votes transfer to $a$, who then has seven votes, one more than $b$. Hence, $a$ wins the election on $V_1$ under the STV rule.

Let $V_2$ contain three votes $b \succ a \succ c$, four votes $a \succ c \succ b$, and six votes $c \succ a \succ b$. Thus, $V_2$ has the same votes as $V_1$, except the roles of $b$ and $c$ are switched. Hence $a$ wins the election on $V_2$ under the STV rule.

Now, consider the set of votes $V_1 + V_2$. $b$ and $c$ each receive nine votes, whereas $a$ receives only eight votes. Hence, $a$ drops out first and cannot win. ∎

The remaining rules are all based on pairwise elections. For these, it is useful to consider the *pairwise election graph* between the candidates, in which there is a directed edge from candidate $a$ to candidate $b$ with weight $w$ if $a$ defeats $b$ by $w$ votes in their pairwise election. For example, if the votes are $a \succ b \succ c$ and $b \succ a \succ c$, then the pairwise election graph is:

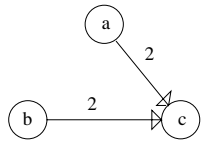

In the remaining proofs, when we apply Lemma 1, it will be easier not to give the votes in $V_1$ and $V_2$, but rather only the pairwise election graphs for $V_1$ and $V_2$. (The pairwise election graph of $V_1 + V_2$ can be inferred from these graphs by summing their edges.) Of course, this approach is legitimate only if we can show that there do indeed exist votes for $V_1$ and $V_2$ that realize these graphs. This is the purpose of the following lemma.

**Lemma 2** *For any pairwise election graph $G$ whose weights are even-valued integers, votes can be constructed that realize $G$.*

**Proof**: To increase the weight on the edge from candidate $a$ to $b$ by 2 without affecting any other weights, we can add the following two votes (where $c_1, c_2, \ldots, c_{m-2}$ are the remaining candidates):

- $a \succ b \succ c_1 \succ c_2 \succ \ldots \succ c_{m-2}$,
- $c_{m-2} \succ c_{m-3} \succ \ldots \succ c_1 \succ a \succ b$.

Hence we can realize any pairwise election graph $G$ with even-valued integer weights. ∎

**Theorem 5** *The Copeland rule is neither an MLEWIV nor an MLERIV rule.*

**Proof**: We will apply Lemma 1 to both cases. Let $V_1$ realize the following pairwise election graph (by Lemma 2):

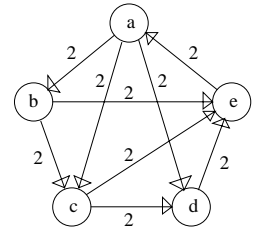

A candidate's Copeland score is the number of outgoing edges minus the number of incoming edges. Therefore, the scores in this election are as follows: $a$ gets 2 points, $b$ gets 1 point, $c$ gets 0 points, $d$ gets $-1$ point, and $e$ gets $-2$ points. Hence the ranking produced by the Copeland rule on $V_1$ is $a \succ b \succ c \succ d \succ e$.

Let $V_2$ realize the following pairwise election graph (by Lemma 2):

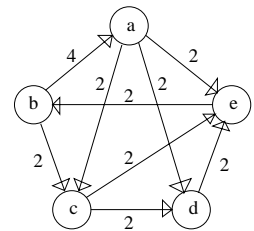

The Copeland scores in this election are as follows: $a$ gets 2 points, $b$ gets 1 point, $c$ gets 0 points, $d$ gets $-1$ point, and $e$ gets $-2$ points. Hence the ranking produced by the Copeland rule on $V_2$ is $a \succ b \succ c \succ d \succ e$, the same as on $V_1$.

The pairwise election graph of $V_1 + V_2$ is the following:

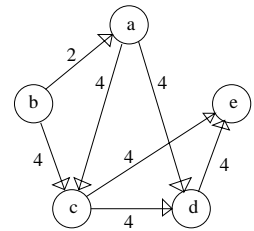

The Copeland scores in this election are as follows: $b$ gets 2 points, $a$ gets 1 point, $c$ gets 0 points, $d$ gets $-1$ point, and $e$ gets $-2$ points. Hence the ranking produced by the Copeland rule on $V_1 + V_2$ is $b \succ a \succ c \succ d \succ e$. ∎

**Theorem 6** *The maximin rule is neither an MLEWIV nor an MLERIV rule.*

**Proof**: We will apply Lemma 1 to both cases. Let $V_1$ realize the following pairwise election graph (by Lemma 2):

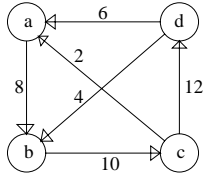

In this election, $a$'s worst pairwise defeat is by 6 votes; $b$'s worst pairwise defeat is by 8 votes; $c$'s worst pairwise defeat is by 10 votes; and $d$'s worst pairwise defeat is by 12 votes. Hence the ranking produced by the maximin rule on $V_1$ is $a \succ b \succ c \succ d$.

Let $V_2$ realize the following pairwise election graph (by Lemma 2):

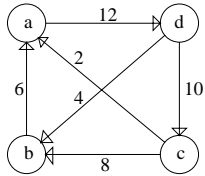

In this election, again, $a$'s worst pairwise defeat is by 6 votes; $b$'s worst pairwise defeat is by 8 votes; $c$'s worst pairwise defeat is by 10 votes; and $d$'s worst pairwise defeat is by 12 votes. Hence the ranking produced by the maximin rule on $V_2$ is $a \succ b \succ c \succ d$, the same as on $V_1$.

The pairwise election graph of $V_1 + V_2$ is the following:

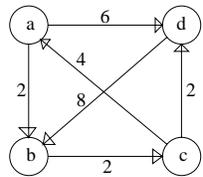

In this election, $c$'s worst pairwise defeat is by 2 votes; $a$'s worst pairwise defeat is by 4 votes; $d$'s worst pairwise defeat is by 6 votes; and $b$'s worst pairwise defeat is by 8 votes. Hence the ranking produced by the maximin rule on $V_1 + V_2$ is $c \succ a \succ d \succ b$. ∎

**Theorem 7** *The ranked pairs rule is neither an MLEWIV nor an MLERIV rule.*

**Proof**: We will apply Lemma 1 to both cases. Let $V_1$ realize the following pairwise election graph (by Lemma 2):

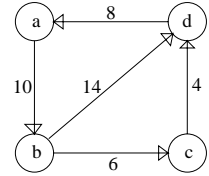

In this election, the pairwise rankings are locked in as follows. First, $b \succ d$ is locked in; then, $a \succ b$ is locked in; then $d \succ a$ is inconsistent with the rankings locked in already, so the next pairwise ranking locked in is $b \succ c$; and finally, $c \succ d$ is locked in. Hence the ranking produced by the ranked pairs rule on $V_1$ is $a \succ b \succ c \succ d$.

Let $V_2$ realize the following pairwise election graph (by Lemma 2):

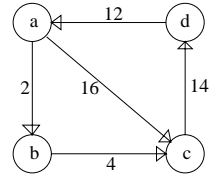

In this election, the pairwise rankings are locked in as follows. First, $a \succ c$ is locked in; then, $c \succ d$ is locked in; then $d \succ a$ is inconsistent with the rankings locked in already, so the next pairwise ranking locked in is $b \succ c$; and finally, $a \succ b$ is locked in. Hence the ranking produced by the ranked pairs rule on $V_2$ is $a \succ b \succ c \succ d$, the same as on $V_1$.

The pairwise election graph of $V_1 + V_2$ is the following:

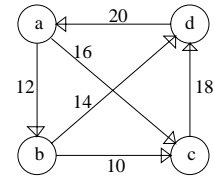

In this election, the pairwise rankings are locked in as follows. First, $d \succ a$ is locked in; then, $c \succ d$ is locked in; then, $a \succ c$ is inconsistent with the rankings locked in already, so the next pairwise ranking locked in is $b \succ d$; then, $a \succ b$ is inconsistent with the rankings locked in already, so the next (and last) pairwise ranking locked in is $b \succ c$. Hence the ranking produced by the ranked pairs rule on $V_1 + V_2$ is $b \succ c \succ d \succ a$. ∎

Among the common voting rules that we have studied, none is an MLEWIV rule but not an MLERIV rule. So, one may wonder if perhaps the MLEWIV property implies the MLERIV property. To see that this is not the case, consider a hybrid rule $A$ that first chooses a winner according to some MLEWIV rule $B$, and produces the ranking of the remaining candidates according to rule $C$ which is not an MLERIV rule. $A$ is an MLEWIV rule because $B$ is, but it is not an MLERIV rule because $C$ is not.

## 5 Conclusions and future research

Voting is a very general method of preference aggregation. We considered the following view of voting: there exists a "correct" outcome (winner/ranking), and each voter's vote corresponds to a noisy perception of this correct outcome. Given the noise model, for any vector of votes, we can compute the maximum likelihood estimate of the correct outcome. This maximum likelihood estimate constitutes a voting rule. In this paper, we asked the following question: *For which common voting rules does there exist a noise model such that the rule is the maximum likelihood estimate for that noise model?* The following table summarizes our results for the rules discussed in this paper.

|  | **MLERIV** | ¬ **MLERIV** |
|---|---|---|
| **MLEWIV** | Scoring rules (incl. plurality, Borda, veto) | Hybrids of MLEWIV and ¬ MLERIV |
| ¬ **MLEWIV** | STV | Bucklin Copeland maximin ranked pairs |

*Classification of voting rules discussed in this paper.*

We believe the techniques that we used to prove these results should be easy to apply to other rules as well.

There are many questions left to be answered by future research, including at least the following. How reasonable are the noise models that we gave to show that certain rules are MLEWIV or MLERIV rules? To the extent that they are not reasonable, can we improve them? Do these improved noise models lead to the same rules, or different (and possibly altogether new) ones? For the rules that we showed are not MLEWIV or MLERIV rules, can we still interpret them as maximum likelihood estimators if we relax somewhat the assumption of votes being drawn independently? (If we have no restriction at all, then Proposition 1 shows that this can always be done.) Alternatively, are there other rules that can be interpreted as maximum likelihood estimators under our independence assumption, while producing outcomes that are "close" to those produced by rules that cannot be so interpreted? Finally, are there rules that are not MLEWIV/MLERIV but for which Lemma 1 cannot be used to show this, or does Lemma 1 in fact also provide us with a *sufficient* condition for a rule being MLEWIV/MLERIV?